\documentclass[12pt]{article}
\usepackage{amsmath,amssymb,graphicx,mathrsfs,hyperref}

\newcommand{\be}{\begin{equation}}
\newcommand{\ee}{\end{equation}}
\newcommand{\bea}{\begin{eqnarray}}
\newcommand{\eea}{\end{eqnarray}}

\def\({\left(} \def\){\right)}

\begin{document}

\title{\vspace{-1.8in}
\vspace{0.3cm} {Universal stress-tensor correlation functions of strongly coupled conformal fluids }}
\author{\large Ram Brustein${}^{(1)}$,  A.J.M. Medved${}^{(2)}$ \\
 \hspace{-1.5in} \vbox{
 \begin{flushleft}
  $^{\textrm{\normalsize
(1)\ Department of Physics, Ben-Gurion University,
    Beer-Sheva 84105, Israel}}$
$^{\textrm{\normalsize (2)  Department of Physics \& Electronics, Rhodes University,
  Grahamstown 6140, South Africa }}$
 \\ \small \hspace{1.7in}
    ramyb@bgu.ac.il,\  j.medved@ru.ac.za
\end{flushleft}
}}
\date{}
\maketitle

\begin{abstract}
We have shown in two accompanying papers that, for  Einstein gravity, the graviton multi-point functions are universal in a particular kinematic region and depend only on the (generalized) Mandelstam variable s.  The effects of the leading corrections to Einstein gravity were shown to be similarly universal,  inducing  a specific difference in the angular dependence. Here we show, relying on the gauge-gravity duality, that the stress-tensor correlation functions of conformal fluids whose gravitational dual is either Einstein gravity or its leading correction are also universal. We discuss the possible significance of these results to multi-particle correlations in heavy-ion collisions. We show that, to test our ideas, the stress-energy correlation functions have to measured rather than the standard multiplicity correlation functions. We then discuss schematically how stress-energy correlations in heavy-ion collisions can be used to test our findings.  We argue that, if these correlations can be measured  precisely enough, they  will provide a unique way to probe the existence of a gravitational dual to the quark-gluon plasma and to determine its universality class.

\end{abstract}
\newpage

\section{Introduction}

The gauge--gravity duality states that strongly coupled gauge theories in four dimensions have a dual description in terms of  weakly
coupled gravity theories in  five-dimensional anti-de Sitter (5D AdS) space \cite{Maldacena1,Maldacena}.
In some sense, the gauge theory lives at the outer boundary of
the AdS spacetime, and so this is  a holographic  correspondence \cite{Witten}.

If the bulk gravity theory is Einstein's,  then all of its on-shell amplitudes depend on a single dimensionful parameter, the 5D Newton's constant $G_5$.
Hence, appropriately chosen ratios of amplitudes  do not depend on $G_5$. For example,
the ratio of the shear viscosity to the entropy density, $\;\eta/s=1/4\pi\;$ \cite{PSS}, is related to a ratio of two-point functions of gravitons
\cite{BMold} and thus independent of $G_5$. In string theory, the leading corrections to Einstein gravity are universal
and depend on one additional dimensionful parameter, the string tension $\alpha'$.

So far, the emphasis in theoretical and experimental studies has been on the two-point functions; the shear viscosity, conductivity, {\it etcetera}.  From the experimental side, the prototypical example of a  strongly coupled fluid is the quark--gluon plasma (QGP) \cite{QCD}. The QGP is produced in heavy-ion collisions  in the Large Hadron Collider (LHC) at CERN and the Relativistic Heavy Ion Collider (RHIC) at Brookhaven. Experiments on the QGP have already had significant success at testing the duality. In particular, the unusually low value of $\eta/s$  was first predicted via the gauge--gravity duality
\cite{PSS}
and then later substantiated \cite{KSS,confirm}.

As multi-particle correlations in heavy-ion collisions have recently been measured \cite{ALICE:2011ab,ALICE:2011ac},  the need for extracting information that can be utilized  to probe the gravitational dual of the QGP has become pertinent. In the context of the gauge--gravity duality, the simplest field-theory objects are the connected correlation functions for the stress-energy tensor. These are dual to graviton $n$-point functions \cite{Witten}.

The main idea that we wish to present in this paper is that, to probe the gravitational dual to the QGP, one should look at stress-energy correlation functions rather than at the standard multiplicity correlation functions.

We have recently calculated the one-particle irreducible (1PI) on-shell amplitudes of gravitons \cite{cavepeeps} and extended the calculation to many connected $n$-point functions \cite{newby}. This work provides the basic ingredients to evaluate the connected stress-energy correlations. We have focused on tensor-graviton amplitudes for a 5D AdS black brane background and on a certain kinematic region in which the amplitudes simplify significantly. The calculation was performed for Einstein gravity and for its leading-order corrected theory, Gauss--Bonnet (GB) gravity. Because the work presented in \cite{cavepeeps,newby} is highly technical and not accessible to theoretical experts in heavy-ion physics or to experimenters, we have made included a short review of the results in a more accessible form.

\section{Review of graviton multi-point amplitudes}

The gravitational Lagrangian is expanded on the background of the AdS black brane whose space part has the topology of a sphere. The expansion is in the number of gravitons, $h_{\mu\nu}$ --- a small perturbation of the metric from its background value, $\;g_{\mu\nu}\to g_{\mu\nu} +h_{\mu\nu}\;$.  The $n^{\rm th}$ order of the  expansion can be used to read off the graviton 1PI $n$-point function.  The bulk $n$-point functions are evaluated at some large but finite radius in preparation for a process of holographic renormalization.

In practice, one chooses the following ansatz for
the gravitons:
$\;h_{\mu\nu}= \phi(r)\exp\left[i\omega t-k z\right]\;$.
Here, $z$ indicates the direction of graviton propagation along
the brane (or along any other spacetime slice
at constant radius) and $r$ is the usual AdS radial coordinate,
ranging  from $\;r=r_h\;$ at the horizon of the black brane
to infinity at the AdS boundary. The radius of the AdS curvature is $L$ and the radius of the boundary sphere is also $R\sim L$ \cite{Witten2}. The remaining two spacelike
directions will be denoted by $x$ and $y$. For future reference,
$\;a,b,\cdots=\left\{t,x,y,z\right\}\;$.

In the radial gauge, for which $\;h_{rr}=h_{ar}=0\;$, the gravitons separate into three distinct classes \cite{PSS2}: tensors, vectors and scalars. For our kinematic region of interest (defined below), amplitudes involving  vector modes on the external lines can not be used to discriminate between different theories and scalars on the external lines can be completely discarded \cite{cavepeeps}.  Thus, we will only be interested in the tensors $h_{xy}$. The results are presented below for  Einstein gravity  in units for which
$1/16\pi G_5$ is set to unity and
$\;
s\;=\;-\frac{1}{2n(2n-1)}\sum_{i=1}^{2n}\sum_{\overset{j=1}{j\neq i}}^{2n} k_i^{\mu}k_{j\mu}\;
$
is a generalization of the standard Mandelstam variable symmetrized over $n$ gravitons.

Considerations are restricted to  a certain
kinematic regime, which we refer to as  ``high momentum",
where the correlation functions of the stress tensor simplify considerably. Also,  the comparison between the leading Einstein result and possible corrections becomes simpler, as the number of derivatives in the interaction vertices is emphasized. In this kinematic regime,  $\;k^2$, $\omega^2$, $s\gg 1/L^2\;$,
where $L$ is the AdS curvature scale. However, because our interest is in fluid hydrodynamics, we take $\sqrt{s}\ll T$, where $T$ is the Hawking temperature temperature. So,
\be
1\;\ll\;\sqrt{s} L \;\ll\; \pi LT\;.
\ee
This is self-consistent, as the validity of the
gauge--gravity duality requires that  $\;\pi LT\gg 1\;$.
We will discuss later the meaning of the high-momentum region from the
boundary-theory perspective.

The simplest example is the two-point function, $\;\lim\limits_{r\to\infty}\langle h_{xy}(k)h_{xy}(-k) \rangle_E\;=\; \left(\frac{L}{r}
\right)^{3}\; k^2\;$. And, in the high-momentum region, the higher-point functions for $\;2n=4,6,8\dots\;$ are given by
$\;
\lim\limits_{r\to\infty}\langle(h_{xy})^{2n}\rangle^E_{1PI} \;=\; A_{2n}
\left(\frac{L}{r}
\right)^{4n-1}\; s\;,
$
with all the odd-point functions vanishing. Here, $\;A_{2n}=\frac{(2n-1) \Gamma\left[n+\frac{1}{2}\right]}{\sqrt{\pi}
\Gamma[n-1]}\;$.

In the context of the gauge--gravity duality, the leading correction to on-shell amplitudes comes from
four-derivative corrections to the Einstein Lagrangian.
It depends
on the coefficient of the Riemann-tensor-squared term \cite{tH,POL,cavepeeps}. Since our interest is in unitary theories whose equations of motion are at most second order in time derivatives, we put the corrections in the GB form.

The dimensionless parameter that measures the relative strength of the GB corrections compared to the leading Einstein result is $\;\epsilon\sim l^2_s/L^2\ll 1\;$, where $l_s$ is the string length.  It also appears in the ratio of the shear viscosity to the entropy density
$ \;
\frac{\eta}{s} =  \frac{1}{4\pi}\left[1-8\epsilon\right]\;
$\cite{BLMSY-0712.0805,same_day}. A term in the bulk Lagrangian with $2(m+1)$ derivatives
scales as $\epsilon^m$. From the boundary-theory point of view, $\;\epsilon \sim \lambda^{-1/2}\;$, where $\lambda$ is the field-theory 't Hooft coupling.

The real distinction between Einstein and GB gravity comes from the non-linear interaction terms. Einstein gravity has
only two-derivative vertices \cite{Hof,cavepeeps}. In contrast, the  four-point and $\epsilon$-order higher-point functions of GB gravity have four derivatives and,
at higher orders in $\epsilon$, there  can be many more. This distinction between different powers of $\epsilon$
is illustrated in Fig.~1.

We have found in \cite{newby} that, to first order in $\epsilon$ and
for $\;2n=4,6,8\dots\;$,
\be
\lim_{r\to\infty}\langle h_2^{2n}\rangle_E \;=\;
\frac{(2n-1) \Gamma\left[n+\frac{1}{2}\right]}{\sqrt{\pi}
\Gamma[n-1]}\left(\frac{L}{r}
\right)^{4n-1} s\;,
\label{eqn2}
\ee
\be
\lim_{r\to\infty}\langle h_2^{2n} \rangle_{GB} =
\lim_{r\to\infty} \langle h_2^{2n} \rangle_{E}
+
\frac{2}{5}\epsilon\dbinom{2n}{4} \frac{\Gamma\left[n+\frac{3}{2}\right]}{\sqrt{\pi}
\Gamma[n-1]}\left(\frac{L}{r}
\right)^{4n+1} s(s+v)\;,
\label{GGen}
\ee
with the understanding that $\;v=0\;$ for $\;2n=4\;$.

The external gravitons are symmetrized in our expressions, with
the center-of-mass variable $s$ and the generalized Mandelstam variable  $v$ accounting for the symmetrization:
\be
s\;=\;-\frac{1}{2n(2n-1)}\sum_{i=1}^{2n}\sum_{\overset{j=1}{j\neq i}}^{2n} k_i^{\mu}k_{j\mu}\;,
\label{mandel}
\ee
\be
v\;=\;-\frac{1}{2n(2n-1)(2n-2)(2n-3)}\sum_{i_1=1}^{2n}\sum_{\overset{i_2=1}{i_2\neq i_1}}^{2n}
\sum_{\overset{i_3=1}{i_3\neq i_{1,2}}}^{2n}\sum_{\overset{i_4=1}{i_4\neq i_{1,2,3}}}^{2n}
\sum_{\overset{j=1}{j\neq i_{1,2,3,4}}}^{2n}
 k_{i_1}^{\mu}k_{j\mu}\;.
\ee

We have also found that, at order $\epsilon$, the connected functions are equal to the 1PI ones up to at least $\;2n=6\;$. Because the $\epsilon$-order theory is similarly constrained by general covariance and a strict number of derivatives, we believe that  this agreement persists for all values of $n$. In this case, the single dimensionful parameter that fixes the form of the amplitudes is the coefficient of the Riemann-tensor-squared term in the Lagrangian.

In an AdS/CFT context, we need  to consider the presence of a source for  scalar gravitons and therefore the possibility that they can propagate in internal lines.  Additionally, scalars that originate from compactifications of string theory can also propagate in internal lines. We have found that, luckily,  the additional scalar contributions are higher order in $\epsilon$. This conforms with the general argument above.

The relevance of the corrections to the interaction terms is determined by the value of $\;L^2 k^2 \epsilon \sim k^2 l_s^2\;$, whereas the consistency of treating string theory effectively as a theory of gravity requires that $\;k^2 l_s^2<1\;$.  So that,
if the momenta are ``stringy" or ``Planckian", then the corrections become substantial. But, as the hydrodynamic approximation requires that $\;L^2 k^2 \epsilon\ll (\pi L T)^2\epsilon \;$, it is really the value of
$\;(\pi L T)^2\epsilon\sim (\pi R T)^2\epsilon\; $  which determines whether the higher-point functions will be substantially corrected. If $\;(\pi R T)^2\epsilon\ll 1 \;$, then the corrections will not appear in higher-point functions, only in the two-point function. This will be important when we discuss the possibility of experimentally determining the corrections.

\section{Stress-energy correlation functions}

We wish to explain how the stress-energy correlation functions are calculated.

The connected bulk Feynmann diagrams will have counterpart Witten diagrams on the boundary (see Fig.~1). The calculation of these Witten diagrams is presented here and  a discussion on holographic renormalization is also provided.

The dual to a bulk graviton $h_{ab}$ is the stress (energy-momentum) tensor $T_{ab}$ of the boundary theory. This follows from the standard bulk--boundary dictionary \cite{Witten,Maldacena} and  the boundary stress tensor (when expressed in gravitational terms) being canonically conjugate to the gravitons \cite{BY,BK}. Hence, it can be expected that a graviton $n$-point function is telling us about a field-theory correlation function with $n$ insertions of a stress tensor. That is,
\be
\lim_{r\to\infty}\langle h_{a_1b_1}h_{a_2b_2}\dots h_{a_nb_n} \rangle_{Con} \;\leftrightarrow\;
\langle T_{a_1b_1}T_{a_2b_2}\dots T_{a_nb_n} \rangle_{Con}\;,
\ee
where the double-sided arrow indicates that the quantities
are  dually related (this is {\em not} an equivalence) and the
subscripts of $Con$ reminds us that the relation is between connected
functions.

To put this on a formal level, we need to apply the standard rules of holographic renormalization \cite{dBVV,skenderis1,skenderis2} to the bulk $n$-point functions. This is a three-step procedure. The first step is to extrapolate the bulk quantity to the boundary. Technically, this requires evaluating at some large but finite value of radius $\;r=r_{0}\;$ and then imposing the limit $\;r_0\to\infty\;$ at the end. This step was carried out  in \cite{newby} and reviewed in
Section~2. The second step is to multiply the result of the first step  by $\Omega^q$,  where $\Omega$ is an appropriate conformal factor and the power $q$ is determined by the conformal dimension of the operators that are being  calculated.  The third step is to subtract off any divergences, since these should correspond to the background contributions. One then only retains the finite part that survives the three steps.

Let us elaborate on the second step. The conformal factor $\Omega$ can be deduced from the asymptotic form of the metric.  In our case of an AdS brane geometry,
\bea
\lim_{r\to\infty} ds^2\;&=&\; -\frac{r^2}{L^2}dt^2 +\frac{L^2}{r^2}dr^2 +\frac{r^2}{L^2}\left[
dx^2 +dy^2 + dz^2\right]\; \nonumber \\
\;&=&\;-\frac{r^2}{L^2}\left[\eta_{ab}dx^{a}dx^{b}\right]+\frac{L^2}{r^2}dr^2\;,
\label{asymetric}
\eea
with the square brackets in the lower line corresponding to four-dimensional flat or Minkowski space (as follows from the Poincar\'e invariance of the boundary). The appropriate conformal factor can now be  identified as $\;\Omega=r/L\;$. One then  multiplies an operator of conformal dimension $\Delta$ by a factor $\Omega^q$ such that $\;q=\Delta-3\;$.  The subtraction of 3 takes into account the contribution of the metric determinant. If we calculate a product of several operators with each operator of (mass) conformal dimension $\Delta_i$, then
 $\;\Delta=\sum_i \Delta_i\;$.

The operators in our case are products of gravitons and graviton derivatives.  Each derivative has dimension $\;\Delta_{\nabla}=1\;$ and each graviton has  $\;\Delta_h=2\;$. The latter can be deduced from the boundary behavior of the metric; in particular, a rescaling of  $\;r\to\alpha r\;$ requires that the metric at the boundary (and, hence, the  gravitons)  be rescaled by $\;g_{ij}\to \alpha^{-2} g_{ij}\;$.

\begin{figure}[t]
\vspace{-1.6in}
\scalebox{.45}{\includegraphics[angle=0]{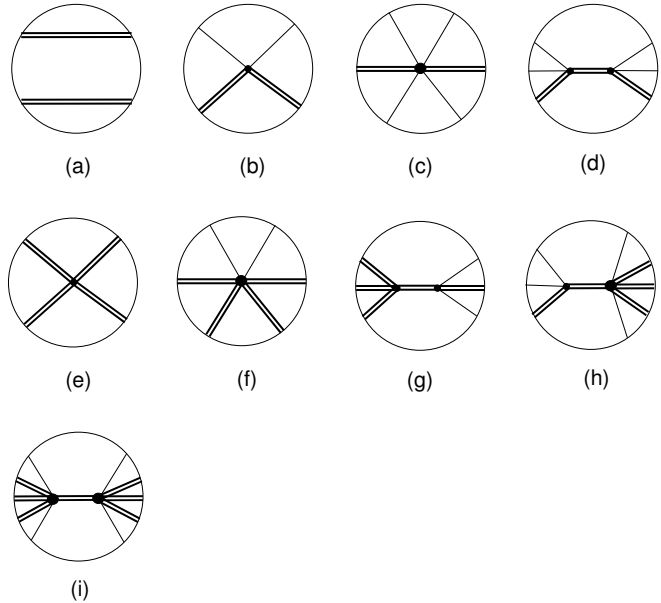}}
\caption{Witten diagrams for stress-tensor correlation functions. The circle represents the AdS outer boundary, the dots depict
interaction vertices, single lines correspond to undifferentiated gravitons and double lines correspond to differentiated gravitons. (a) A disconnected four-point
function. (b) A 1PI four-point function at order $\epsilon^0$. (c) A 1PI six-point function at order $\epsilon^0$. (d) A 1PR connected six-point function
at order $\epsilon^0$.  (e) A 1PI four-point function at order $\epsilon$.  (f) A 1PI six-point function at order $\epsilon$. (g) A 1PR connected six-point function at order $\epsilon$. (h) A 1PR connected eight-point function at order $\epsilon$. (i) A 1PR connected ten-point function at order $\epsilon^2$.}
\end{figure}

For instance, let us consider the four-point connected function at order $\epsilon^0$. At this order, the only contribution comes from the 1PI four-point function that is depicted in diagram (b) of Fig.~1. Since there are four gravitons and
two derivatives, $\;\Delta=4\cdot2+2=10\;$ and $\;q=10-3=7\;$.
We then have the renormalized function ({\em cf}, Eq.~(\ref{eqn2}) with
$\;n=2$),
\bea
\left[\langle h_2h_2h_2h_2\rangle_{Con}^E\right]_{Ren} &=&
\lim_{r\to\infty}\left[\Omega^q \langle h_2h_2h_2h_2\rangle_{Con}^E\right]
\nonumber \\
 &=&
\lim_{r\to\infty}\left[\left(\frac{r}{L}\right)^7
\frac{3 \Gamma\left[\frac{5}{2}\right]}{\sqrt{\pi}
\Gamma[1]}
\left(\frac{L}{r}
\right)^7s\right]\;+\;\cdots \nonumber \\
&=&
\frac{9}{4} s\;,
\label{4con}
\eea
where the ellipsis in the middle line  stands for subdominant  contributions that vanish when the final limit is taken. From Eq.~(\ref{4con}), we can read off the angular dependence and numerical coefficient of  the four-point correlation function of the stress-tensor, $\langle T_{a_1b_1}T_{a_2b_2}T_{a_3b_3} T_{a_4b_4} \rangle_{Con}$.

The order-$\epsilon$ correction to the four-point function is depicted in diagram (e) of Fig.~1. We evaluate it in a similar manner, except that the extra factor of $L^2/r^2$ in Eq.~(\ref{GGen}) is exactly compensated by  $q$ increasing from $7$ to $9$ on account of the  two extra derivatives at this order.

Following this described procedure, we find that all powers of $r$ and $L$ are stripped away from the bulk expressions, just as in the the examples above. What is left is a quantity that is finite and well defined at the boundary. Hence, the third step in the holographic renormalization procedure turns out not to change the result. This can be understood by realizing that the background is never an issue in the high-momentum region, and subleading contributions to the metric could be important in principle but, in our case, only show up in terms that are asymptotically vanishing. It is worth emphasizing that the simplicity of the process is, in our case, a consequence of working in the high-momentum regime. The point is that, in this kinematic region, the only relevant derivatives are with respect to $z$ and $t$, and it so happens that $g^{zz}$, $g^{tt}$ are guaranteed to have the same radial structure and to be  dispersed democratically at the boundary of the AdS spacetime. For a general kinematic region, the procedure will  in general be  technically more involved.

The end result is the following expressions for the Einstein and Gauss--Bonnet-corrected two-, four- and six-point correlation functions (for all of these, it is implied that the numbers labeling the stress tensors have been fully symmetrized):
\be
\langle T_{xy}T_{xy} \rangle_{Con}^E\;=\;  k_1^2\;
T_{xy}^{(1)} T^{(2)}_{xy}\;,
\ee
\be
\langle T_{xy}T_{xy}T_{xy}T_{xy}\rangle_{Con}^E\;=\;
\frac{9}{4}
\;s\;T_{xy}^{(1)}
T^{(2)}_{xy} T_{xy}^{(3)}T_{xy}^{(4)}\; ,
\ee
\be
\langle(T_{xy})^{6}\rangle^E_{Con}\;=\;
\frac{75}{8}\;s\;
T_{xy}^{(1)}
T^{(2)}_{xy} T_{xy}^{(3)} T_{xy}^{(4)}
T_{xy}^{(5)}T_{xy}^{(6)}\; ,
\ee
\be
\langle T_{xy}T_{xy} \rangle^{GB}_{Con}\;=\; \left[1-8\epsilon\right]
 k^2_1\;
T_{xy}^{(1)} T^{(2)}_{xy}\;,
\ee
\be
\langle T_{xy} T_{xy}T_{xy}T_{xy}\rangle^{GB}_{Con}\;=\;
\frac{3}{4}\;
\left[3s+\epsilon\; s^2\right]\;T_{xy}^{(1)}
T^{(2)}_{xy} T_{xy}^{(3)}T_{xy}^{(4)}\; ,
\ee
\be
\langle (T_{xy})^6 \rangle^{GB}_{Con}\;=\;
\frac{15}{8}\;\left[5s+ 21 \epsilon\; s(s+v)\right]\;
T_{xy}^{(1)}
T^{(2)}_{xy} T_{xy}^{(3)} T_{xy}^{(4)}
T_{xy}^{(5)}T_{xy}^{(6)}\; ,
\ee
whereas  the odd-numbered correlators are trivially vanishing.
The Witten diagrams that correspond to Eqs.~(10-14) are depicted in Fig.~1, diagrams (b)-(g).

We observe that the field-theory correlation functions
exhibit the very same angular dependence
as their  bulk correspondents.
But one might wonder how these results would be corrected in the event
of six- and   higher-derivative Lovelock terms.
Two more derivatives means another factor of order $\epsilon$.
Then, as will be
made clear in the discussion below, any such correction is
is suppressed by at least  $N^{-1/2}<<1$, where
$N$ is the ``number of colors''.

Note that we always presume truncated correlators in momentum space;
meaning that all external momenta are stripped away
except those due to the explicit derivative operators in the original action
(the internal momenta are, of course, integrated over).
Hence, the correlation functions are really just numbers which can be
predicted from either the gauge theory (at least in principle) or
from its gravitational dual.

It is instructive to understand the kinematics of the high-momentum regime from the boundary point of view. Recall that
we consider geometries in which the boundary theory is defined on a sphere
$S^3$, with the radius $R$ of the sphere scaling as the AdS scale  $\;R\sim L\;$. So, in the boundary theory, an extra dimensionful parameter is introduced which breaks conformal invariance spontaneously in the same way that the temperature does. In this context, the quantity  $\;\pi T L\sim \pi T R\;$ corresponds to the ratio of  the size of the boundary fluid to the thermal wavelength.  The radius $R$ can be identified with the
spatial extent of the fluid \cite{Witten2,Gubser}.
The high-momentum condition  is then $\;kR\gg 1\;$ and,
since we are also interested in the hydrodynamic limit, $\;k/\pi T\ll 1\;$, where $T$ is now the fluid temperature,
\begin{equation}
\;1\ll \sqrt{s}R\ll \pi T  \;.
\end{equation}

\section{Contact with experiment}

We now wish to translate the previous findings into analogous statements about the QGP. The current objective is then to provide  a schematic description of how the stress-tensor correlation functions of  the previous section might be experimentally tested.

The multi-point correlation functions of the stress-energy tensor are the key to testing our ideas.  These correlators are the most accessible  correlations functions from the AdS/CFT point of view. The reasoning  is that, on the gravity side of the duality, the graviton correlations are the fundamental objects that can be directly calculated, and these correspond to correlations of the stress-energy tensor of the gauge theory. In contrast, the standard tool in heavy-ion scattering  experiments is rather multiplicity correlation functions. This distinction ({\em i.e.}, energy rather than  multiplicity) will be central to the following discussion.

Measuring and calculating accurate multi-particle correlations and then comparing these results to theory in an effective way is essential
if one is to determine  $\eta/s$ and, even more importantly, determine whether the QGP in ALICE (and other heavy-ion experiments) has a gravity dual. The reason that multi-particle correlations are important is because they provide additional constraints on any  theory that is supposed to predict $\eta/s$. For example, if the dual theory is Einstein gravity, not only is the ratio $\eta/s$ equal to  $1/4\pi$
but  all the higher-point correlations are  also completely fixed numbers (a zero-parameter theory). A theory that allows deviations from $\;\eta/s = 1/4\pi\;$ also comes equipped with specific modifications for the multi-particle correlations.

To begin, let us imagine a heavy-ion collision that creates a drop of QGP fluid. The drop expands and cools and then, after it freezes, a variety of decay products (mostly pions) stream outward. The energy, momentum and the velocity of the decay products can be measured (more easily for charged particles than for neutral ones), so that it is possible to measure the angular distribution of energy, momentum and velocitiy for a large number of particles emanating from the QGP fluid. It should then be possible to deduce the corresponding quantities  of the fluid well before freeze out. These should be used to determine the correlation functions of the stress-tensor components  in a way similar to the standard methods of  evaluating multiplicity correlation functions ({\it e.g.}, \cite{MCF}).

Let us now consider the relevant components of the stress tensor $\;T_{xy}=T_{yx}\;$; that is, the duals to the tensor graviton modes. We are  assuming, without
loss of generality, that the fluid flows in the  $z$ direction.
For a conformal boundary theory, the stress tensor can be expressed to leading order as ({\it e.g.} \cite{SS})
\be
T_{ab}\;=\;\frac{\rho}{3}\Bigg[4u_au_b+\eta_{ab}
-4\pi\frac{\eta}{s}\frac {P^c_{\ a}P^d_{\ b}}{\pi T}
\left(\partial_cu_d
+\partial_du_c-\frac{2}{3}\eta_{cd}\eta_{cd}\partial_eu^e\right)
\Bigg]\;,
\label{stress}
\ee
where $u^a$ is the four-velocity of the fluid,
$\;P^{ab}=\eta^{ab}+u^au^b\;$
projects vectors onto directions perpendicular to $u^a$,
$\rho$ is the energy density, $T$ is the temperature
and  $\eta/s$ is the ratio of shear viscosity to
entropy density. To arrive at this form, we have used the relations
 $\;p=\rho/3\;$  for the pressure and $\;s=(\rho+p)/T=(4/3)\rho/T\;$.
Indices are raised or lowered with the Minkowski metric $\eta_{ab}$.

We next define
$\;4\pi\frac{\eta}{s} \Phi\equiv  T^x_{\ y}$ and
a polarization tensor $\widehat{\epsilon}^{cd}\equiv\frac{1}{2}\epsilon^{ab}P^c_{\ a}P^d_{\ b}\;$ in terms of the  ``binormal vector''
 $\epsilon^a_{\ b}\equiv\left\{1\;{\rm if}\;
a\neq b\; ; \; 0\;{\rm if}\; a=b\;\right\}$.
Then we have
\be
\Phi\;=\; -\widehat{\epsilon}^{ab}
\frac{\rho}{3}
\left(\tilde\partial_{ a}u_b
+\tilde\partial_{ b}u_a\right)\;,
\label{phidef}
\ee
where $\tilde{\partial}_{ a}$ is  a dimensionless spatial derivative, measured in units of the temperature $\;\tilde{\partial}=\frac{1}{\pi T}\partial\;$.  $\Phi$
can be  determined experimentally  by measuring the angular distribution of the energies, momenta and velocities of the collision products.

For measuring $\Phi$, one needs to reconstruct the energy density of the fluid and its velocity field. This requires measuring the mass and momentum of the decay particles and then using a hydrocode to reconstruct the corresponding values for the fluid. Since this procedure has never been attempted, it is currently unknown  what accuracy can be achieved. The first step in trying to estimate this accuracy would be to try to reconstruct the energy density of the fluid  $\rho$ and its local velocity $u^a$. Then, it will be possible to estimate  the precision in which this reconstruction can be done.

The information about the energy and momentum components is available for charged particles. It is unclear whether the same information will be available for neutral particles. In any event, if charged particles are a good representative of the overall distribution, then this would be sufficient for our purposes.

The main expected difficulty is how to handle the derivatives of the four-velocities that appear in the expression for $\Phi$ in Eq.~(\ref{phidef}). However, such a derivative can be evaluated in Fourier space (see below)  by multiplying the velocity with its associated momentum. This is a procedure that can be expected to be performed rather accurately.

To further elaborate,  one would determine  the contribution $\Phi_i(t_i,\vec{x_i})$ corresponding to a ``fluid particle" $i$  at time $t_i$ in position $\vec{x}_i$. Incorporating all the available data points into a single distribution, one would end up with $\Phi$  as a function of space-time coordinates. As our previous expressions are in momentum space, it then becomes the ``simple'' matter of performing  an $n$-fold Fourier transform on a product of $n$ $\Phi$'s. For instance, the four-point correlation function would necessitate the transformation
 \\
$\langle \Phi(p^{\mu}_1) \Phi(p^{\mu}_2) \Phi(p^{\mu}_3) \Phi(p^{\mu}_4)
\rangle \;=\;$
\be
 \int d^4x_1 \int d^4x_2 \int d^4x_3 \int d^4x_4
e^{i(p_1\cdot x_1 +p_2\cdot x_2 + p_3\cdot x_3 +p_4\cdot x^4)}
\Phi(x^{\mu}_1)\Phi(x^{\mu}_2)\Phi(x^{\mu}_3)\Phi(x^{\mu}_4)\;,
\ee
where we have used the same symbol but a  different argument to denote
the Fourier-transformed quantity. This is similar to the procedure for evaluating multiplicity correlation functions.

The quantities that we  have used the gauge--gravity duality to calculate  are related to energy correlations in a thermal  state $|\Psi_{ther}\rangle$,
\be
\frac {\langle \Psi_{ther}|\Phi_{ij} \Phi_{kl} \dots \Phi_{pq}|\Psi_{ther}\rangle} {\langle \Psi_{ther} |\Psi_{ther}\rangle}\;,
\label{onestar}
\ee
where $\Phi_{ij}$ is a specific component of the stress-energy tensor. Notice that the thermal averages are {\em not} necessarily Gaussian for all components of $\Phi_{ij}$ (they may be for some) and their connected $n$-point correlations will obey very specific relations among themselves.

The measured quantities  (if they can be measured, as  previously discussed)  would be
\be
\frac{\langle E_{i^{\prime}}| \Phi_{ij} \Phi_{kl} ... \Phi_{pq} |E_{i^{\prime}} \rangle}
{\langle E_{i^{\prime}}|E_{i^{\prime}}\rangle}\;,
\label{twostars}
\ee
where $|E_{i^{\prime}}\rangle$ is the specific state that is created in the experiment in a single collision. Any given state $|E_{i^{\prime}}\rangle$ is expected to be rather similar to a thermal state but not exactly. Then, it is possible to average the correlations over  $n^{\prime}$ different collision events, $\;i^{\prime}=1,\dots n^{\prime}\;$,
\be
\sum_{i^{\prime}=1}^{n^{\prime}}\frac{\langle E_{i^{\prime}}| \Phi_{ij} \Phi_{kl} ... \Phi_{pq} |E_{i^{\prime}}\rangle}
{\langle E_{i^{\prime}}|E_{i^{\prime}}\rangle} \label{threestars}\;.
\ee

To relate the correlations~(\ref{onestar}) and~(\ref{threestars}), one needs to have a good understanding of  the statistical distribution of the created initial states, as well as the statistical description of the evolution (using a hydrocode, for example) of any given initial state to the final state. The key point, though, is that this evolution is a fixed process for all collisions, so that  only the initial state ever changes. In general, one would expect that,  for a  large number of collisions, the averages will be similar to thermal averages. The importance of studying  the statistical properties of the initial states has already
been realized in the context of particle-number correlations; see, {\it e.g.}, \cite{1105.3865} and \cite{MCF}.

Further development of these ideas will require knowledge about
which components of the stress-energy tensor  can be extracted from multi-point correlations in a reasonably accurate way. The best answer is ``all"; meaning, the  energy $T_{00}$, momentum $T_{i0}$ and stress $T_{ij}$.
If these can indeed be measured, one  can calculate their correlations
and compare the results with those of the  experiment.

\section{Scales, couplings and kinematics}

The holographic  dictionary  relates the number of colors $\;N$ to the 't Hooft coupling $\lambda$ and the Yang-Mills coupling $g_{YM}$ of the field theory as $\;\lambda=g_{YM}^2 N\;$. In the large-$N$ limit,  $N\to\infty$, $g_{YM}\to 0$ such that $\lambda$ is finite and large. This means that $1/N$ corrections are neglected while $1/\lambda$ corrections are considered small but finite. But, for the QGP, $\;\lambda \sim 10\;$ while $\;N= 3\;$ and there appears to be a conflict of interests. Nevertheless, as many features of the duality seem to be insensitive to this discrepancy, it is standard to treat $N$ as much larger than $\lambda$  (see \cite{recent} for a recent discussion). We will adhere to this philosophy.

In the hydrodynamic approximation,  momenta and frequencies are expected to be substantially smaller than the QGP temperature $T$. However,
let us adopt the optimistic viewpoint that the hydrodynamic regime can be loosened to
$\;(k/\pi T)^2\;$, $\;(\omega/ \pi T)^2\lesssim 1\;$ and, within the context of our results, $\;s/(\pi T)^2 \lesssim 1\;$.
For the QGP at the LHC,
$\;T\sim 400\;{\rm MeV}\;$ and $\;R\sim 7\;{\rm fm}\;$, so that $\;\pi TR \sim 15\;$ and the range in $s$ is about 200. This means that the opportunity to distinguish different powers of $s$ from data is open.

As discussed previously, the value of the dimensionless parameter $\epsilon (\pi R T)^2$ determines whether the corrections can be substantial in higher-point functions. For the QGP, using the above estimates and $\;\epsilon=1/8 (1-4\pi \eta/s)\;$, we find that $\;\epsilon (\pi R T)^2 \sim 25(1-4\pi \eta/s)\;$. Since $4\pi \eta/s$ is of order unity \cite{Gale:2012rq}, it is likely that the higher-point functions are potentially sensitive to small corrections away from Einstein gravity.

\section{Conclusion}

We have calculated   the  $n$-point
correlation  functions for the stress tensor
of  a strongly coupled conformal fluid via the gauge--gravity duality. We have explained schematically   how to compare these results to statistical tests on the quark--gluon plasma at the LHC and what are the possible outcomes of such a comparison.

Given that the experimental tests are indeed carried out, there are three distinct possibilities: (i) The gravitational dual
is shown, within errors, to be in the universality class of Einstein gravity,
(ii) the gravitational dual is shown to be in the universality class of a corrected Einstein gravity or (iii) the correlation functions of QGP do not follow from either.
Each of the three possibilities would provide very interesting information about the nature of the QGP and about the existence of a gravity dual. In particular, the second would be  a wonderful opportunity to study ``Planckian" physics with the QGP.

\section*{Acknowledgments}

We thank Ofer Aharony, Misha Lublinsky, Sasha Milov, Kyriakos Papadodimas, Jurgen Schukraft,  Sheer El-Showk, Raimond Snellings,  Amos Yarom for helpful discussions. The research of RB was supported by the Israel Science Foundation grant no. 239/10. The research of AJMM received support from
a Rhodes University Discretionary Grant.

\end{document}